\documentclass[aps, pra, twocolumn, tightenlines, footinbib, superscriptaddress,reprint]{revtex4-2}
\usepackage{amsmath}
\usepackage{amssymb}
\usepackage{mathrsfs}
\usepackage{graphicx}
\usepackage{booktabs}  
\usepackage{multirow}  
\usepackage[colorlinks=true, letterpaper=true, pdfstartview=FitV, linkcolor=blue, citecolor=blue, urlcolor=blue]{hyperref}

\begin{document}

\title{Scaling Enhancement of Photon Blockade in Output Fields}

\author{Zhi-Hao Liu}
\affiliation{Key Laboratory of Low-Dimensional Quantum Structures and
Quantum Control of Ministry of Education, Key Laboratory for Matter
Microstructure and Function of Hunan Province, Department of Physics and
Synergetic Innovation Center for Quantum Effects and Applications, Hunan
Normal University, Changsha 410081, China}

\author{Xun-Wei Xu}
\email{xwxu@hunnu.edu.cn}
\affiliation{Key Laboratory of Low-Dimensional Quantum Structures and
Quantum Control of Ministry of Education, Key Laboratory for Matter
Microstructure and Function of Hunan Province, Department of Physics and
Synergetic Innovation Center for Quantum Effects and Applications, Hunan
Normal University, Changsha 410081, China}
\affiliation{Institute of Interdisciplinary Studies, Hunan Normal University, Changsha, 410081, China}

\date{\today}

\begin{abstract}
Photon blockade enhancement is an exciting and promising subject that has been well studied for photons in cavities. However, whether photon blockade can be enhanced in the output fields remains largely unexplored.
We show that photon blockade can be greatly enhanced in the mixing output field of a nonlinear cavity and an auxiliary (linear) cavity, where no direct coupling between the nonlinear and auxiliary cavities is needed.
We uncover a biquadratic scaling relation between the second-order correlation of the photons in the output field and intracavity nonlinear interaction strength, in contrast to a quadratic scaling relation for the photons in a nonlinear cavity.
We identify that this scaling enhancement of photon blockade in the output field is induced by the destructive interference between two of the paths for two photons passing through the two cavities.
We then extend the theory to the experimentally feasible Jaynes-Cummings model consisting of a two-level system strongly coupled to one of the two uncoupled cavities, and also predict a biquadratic scaling law in the mixing output field.
Our proposed scheme is general and can be extended to enhance blockade in other bosonic systems.
\end{abstract}

\maketitle

\section{Introduction}

Single-photon resource~\cite{Aharonovich2016NaPho,Wang2019NaPho,liu2024super} with simultaneous high degrees of efficiency, single-photon purity, and photon indistinguishability, is a crucial device in the implementation of quantum communication~\cite{Pan2012RMP}, quantum computing~\cite{Kok2007RMP}, and quantum metrology~\cite{Muller2017PRL}. Photon blockade, preventing the resonant injection/emission of more than one photon~\cite{Imamoglu1997PRL}, provides an efficient way for single-photon generation with high purity. The field of photon blockade is extended to the high-order~\cite{Miranowicz2013PRA,Hamsen2017PRL,Haider2023PRA,Munoz2014NaPho,Bin2020PhRvL}, multi-mode~\cite{Chakram2022NatPh}, and multi-dimensional~\cite{LiLin2020Sci} correlations, and its application is expanded from generating single photons to demonstrating photonic quantum logic gate~\cite{ShiS2022NatCo,LiM2020PRAPP} and fractional quantum Hall state~\cite{WangC2024Sci}.

Since the prediction of photon blockade~\cite{Imamoglu1997PRL}, diverse efforts have been made to observe and enhance the effect.
Strong coupling between light and matter at the single-photon level enabled the observation of photon blockade in experiments, including single atoms coupled to an optical resonator~\cite{Birnbaum2005Natur,Dayan2008Sci,Aoki2009PRL}, a quantum dot coupled to a photonic crystal resonator~\cite{Faraon2008NatPh,Reinhard2012NaPho}, and a superconducting qubit coupled to a transmission line resonator~\cite{LangC2011PRL,Hoffman2011PRL}.
Different from the photon blockade based on strong nonlinearity (conventional photon blockade), Liew and Savona showed that photon blockade also can be achieved with weak nonlinearity via quantum interference, i.e., unconventional photon
blockade (UPB)~\cite{Liew2010PRL,Bamba2011PRA}. Subsequently, UPB has been predicted in different setups~\cite{Majumdar2012PRL,Xu2013JPB,Xu2014PRA1,Xu2014PRA2,ZhouYH2015PRA,Flayac2017PRA,Zou2020OE}, and has been observed for both optical~\cite{Snijders2018PRL} and microwave~\cite{Vaneph2018PRL} photons.
While the UPB with weak nonlinearity is interesting, the fact that there is a very small amount of photon in the cavity makes it inconvenient for applications~\cite{Lemonde2014PRA}.
Besides, photon blockade is also predicted by nonlinear driving~\cite{Andrew2021sciadv,MaYX2023PRA} and nonlinear loss~\cite{ZhouYH2022PRAPP,SuX2022PRA,Ben-Asher2023PRL,ZuoYL2022PRA}.  Moreover, photon blockade enhancement is proposed based on multimode-resonant interaction~\cite{Stannigel2012PRL,Ludwig2012PRL}, non-Hermitian coupling~\cite{Huang2022LPR,LiJH2015PRA,ZhouYH2018PRA,SunJY2023PRA}, dynamical excitation~\cite{Ghosh2019PRL,LiM2022PRL}, and coupled-resonator chain~\cite{WangY2021PRL,li2023enhancement,lu2024chiral}.

We note that the previous works mainly focus on the photon blockade in the cavities, but photon statistics in the output fields becomes even more complex, such as we can observe photon antibunching for the reflected light but bunching for transmitted light~\cite{Aoki2009PRL,LiaoJQ2010PRA,ShiT2011PRA}. 
Photon statistics for the mixing of two output channels has been investigated in Ref.~\cite{Flayac2013PRA}, and
it shows that the photon antibunching in the mixing output field is not suppressed but rather just displaced in a different region of the system’s parameter space.
Besides, photon antibunching as well as bunching effect are observed in the mixed field of a narrow band two-photon source
and a coherent field~\cite{LuYJ2001PRL}, and tunable photon statistics have been proposed in the admixing of a coherent state
with a squeezed state~\cite{Zubizarreta2020LPRv,Casalengua2020PRA}.
Nevertheless, \emph{whether photon blockade can be enhanced in the mixing fields output from two cavities} hasn't been studied thoroughly. 

In this paper, we combine conventional and unconventional photon blockade and show that \emph{photon blockade can be greatly enhanced in the mixing fields output from a nonlinear cavity and an auxiliary linear cavity}. 
Different from the previous works on UPB in weakly nonlinear photonic molecules~\cite{Liew2010PRL,Bamba2011PRA,Majumdar2012PRL,Xu2013JPB,Xu2014PRA1,Xu2014PRA2,ZhouYH2015PRA,Flayac2017PRA,Zou2020OE}, here the photon blockade enhancement in the output fields is achieved without direct coupling between the two cavities, which brings three advantages. Firstly, there is no time oscillation in the temporal second-order correlation function. Secondly, the single-photon output efficiency is relatively greater. Thirdly, there is no strict relationship between the nonlinear strength and the coupling strength of the two cavities to observe optimal photon blockade.

We analytically identify that there is a \emph{biquadratic scaling relation between the second-order correlation of the photons in the output field and the intracavity nonlinear interaction strength}, in contrast to a quadratic scaling law for the photons in a nonlinear cavity.
Our scheme is general and can be extended to other platforms. As an example, we consider an experimentally feasible Jaynes-Cummings (JC) model for two (uncoupled) cavities with a two-level system (TLS) coupled to one of them, and demonstrate a \emph{biquadratic scaling relation between the second-order correlation of the photons in the output field and TLS-cavity interaction strength}.

\section{$\chi^{(3)}$ model}

\begin{figure}[tbp]
\includegraphics[bb=52 299 536 694, width=8.5 cm, clip]{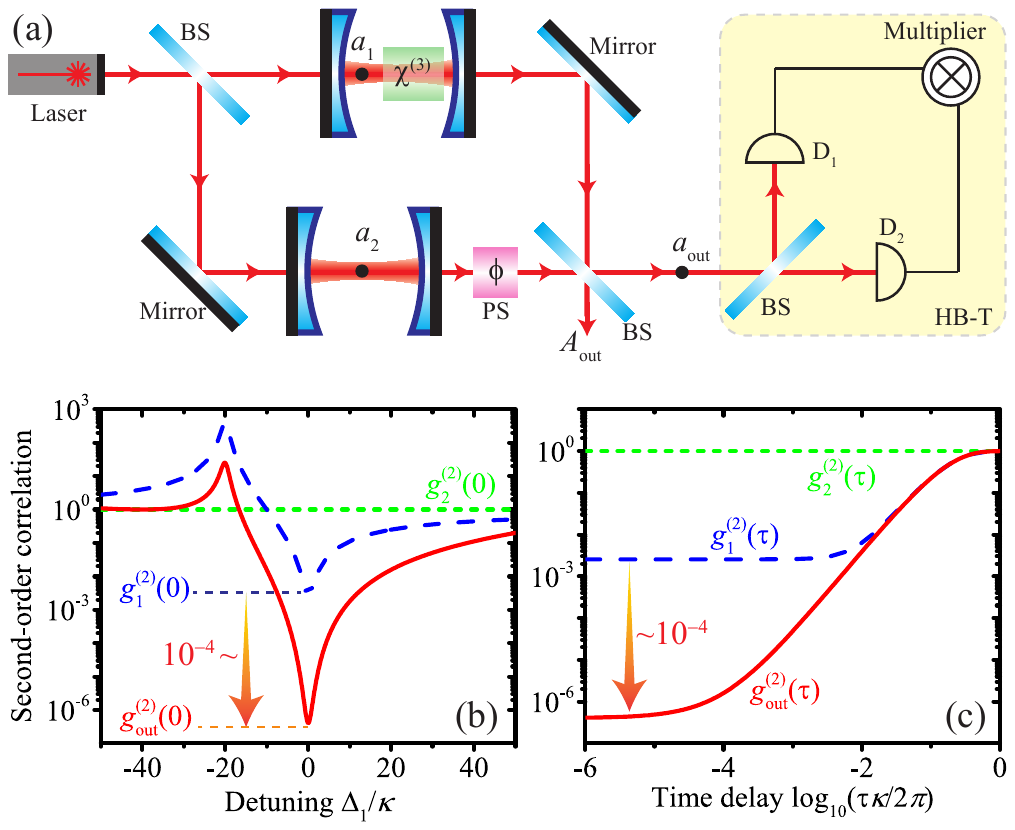}
\caption{(Color online) (a) A Mach–Zehnder interferometer with two cavities ($a_1$ and $a_2$) in the paths. A laser is divided into two beams by a $50/50$ beam splitter (BS), and they are injected into a cavity containing $\chi^{(3)}$ nonlinear medium and an auxiliary (linear) cavity, respectively. The output fields from these two cavities mix by another BS. A phase shifter (PS) is placed in one path to induce tunable phase difference $\phi$ between the two paths. The second-order correlation of the output field is measured by a Hanbury-Brown-Twiss (HB-T) set-up. 
Second-order correlations of the photons in the two cavities [$g^{(2)}_1(\tau)$ and $g^{(2)}_2(\tau)$] and the mixing output field [$g^{(2)}_{\rm out}(\tau)$] are plotted (b) as functions of the detuning $\Delta_1/\kappa$ for the time delay $\tau=0$, and (c) as functions of $\log_{10}(\tau\kappa/2\pi)$ for $\Delta_1=0$.
The parameters are $\phi=\pi$, $U=20\kappa$, and $\delta=2U$.}
\label{fig1}
\end{figure}

Without loss of generality, we first consider the photon blockade in the mixing fields output from a cavity containing $\chi^{(3)}$ nonlinear medium and an auxiliary (linear) cavity [Fig.~\ref{fig1}(a)].
The total Hamiltonian of the system in the frame rotating at the probe laser frequency $\omega_p$ can be written as ($\hbar=1$),
\begin{eqnarray}
H &=&\Delta_1 a_{1}^{\dag }a_{1}+Ua_{1}^{\dag }a_{1}^{\dag
}a_{1}a_{1}+i\varepsilon \left( a_{1}^{\dag }-a_{1}\right) \nonumber \\
&&+ \Delta_2 a_{2}^{\dag }a_{2}+i\varepsilon \left(
a_{2}^{\dag }-a_{2}\right) ,
\end{eqnarray}
where $a_{i}$ and $a^{\dag}_{i}$ are the annihilation and creation operators of the $i$th cavity with frequency $\omega_i$ ($i=1,2$), $\Delta_i=\omega_i-\omega_p$ is the laser detuning from the cavity resonance, $\delta=\omega_2-\omega_1$ is the detuning between the two cavities, $U$ is the nonlinear interaction strength, and $\varepsilon$ is the pumping strength on each cavity.
According to the input-output relation~\cite{Gardiner1985PRA}, the mixing fields $a_{\mathrm{out}}$ and  $A_{\mathrm{out}}$ output from the two cavities can be described by $a_{\mathrm{out}}=(\sqrt{\kappa _{1}}a_{1}+e^{i\phi }\sqrt{\kappa _{2}}a_{2})/\sqrt{2}-a_{\mathrm{vac}}$ and $A_{\mathrm{out}}=(\sqrt{\kappa _{1}}a_{1}-e^{i\phi }\sqrt{\kappa _{2}}a_{2})/\sqrt{2}-a^{\prime}_{\mathrm{vac}}$, where $\kappa_i$ is the one-sided decay rate of the $i$th cavity, $\phi$ is the relative phase between the two output fields (tunable by using the phase shifter), and $a_{\mathrm{vac}}$ ($a^{\prime}_{\mathrm{vac}}$) is the input vacuum field from the right-hand side of the cavities. Here, we focus on the output field $a_{\mathrm{out}}$, and the results of $A_{\mathrm{out}}$ can be obtained just by replacing $\phi$ by $\phi+\pi$ (see Appendix~\ref{AppA}).

Photon statistics of the output field in the steady state can be described by the second-order correlation function
\begin{eqnarray}\label{g_out}
g_{\mathrm{out}}^{\left( 2\right) }\left( \tau\right)  &=&\frac{\left\langle a_{%
\mathrm{out}}^{\dag }a_{\mathrm{out}}^{\dag }\left( \tau\right)a_{\mathrm{out}}\left( \tau\right)a_{\mathrm{out}%
}\right\rangle }{\left\langle a_{\mathrm{out}}^{\dag }a_{\mathrm{out}%
}\right\rangle ^{2}}\\
&=&\sum_{j,k,l,m=1}^{2}e^{i n\phi }\sqrt{\kappa
_{j}\kappa _{k}\kappa _{l}\kappa _{m}}\frac{\left\langle a_{j}^{\dag
}a_{k}^{\dag }\left( \tau\right)a_{l}\left( \tau\right)a_{m}\right\rangle }{N_{\mathrm{out}}^{2}}\nonumber,
\end{eqnarray}%
where $ n= l+m-j-k$, $N_{\mathrm{out}}=\kappa _{1}\langle a_{1}^{\dag }a_{1}\rangle
+\kappa _{2}\langle a_{2}^{\dag }a_{2}\rangle +2\sqrt{\kappa
_{1}\kappa _{2}}{\rm Re}( e^{i\phi }\langle a_{1}^{\dag
}a_{2}\rangle ) $, and $\tau$ is the time delay.
Different from the second-order correlation function in the cavities $g_{i}^{( 2) }( 0) =\langle a_{i}^{\dag}a_{i}^{\dag }a_{i}a_{i}\rangle /\langle a_{i}^{\dag}a_{i}\rangle ^{2}$ ($i=1,2$), $g_{\mathrm{out}}^{(2)}(0)$ also depends on the cross-correlation between
the two cavities (i.e., $\langle a^{\dagger}_2 a_2 a^{\dagger}_1 a_1\rangle$, $\langle a^{\dagger}_1 a^{\dagger}_1 a_2 a_2\rangle$, $\langle a^{\dagger}_1 a^{\dagger}_1 a_1 a_2\rangle$, and $\langle a^{\dagger}_2 a^{\dagger}_1 a_2 a_2\rangle$), and there are phase factors $e^{in\phi }$ in front of the terms,
which can be negative and induce the enhancement of photon blockade in the
output field, without changing the photon statistics in the cavities.

The dynamics of the system are governed by the master equation~\cite{Carmichael1993} $d\rho /dt=-i\left[ H,\rho \right] +\sum_{i=1,2}\kappa
_{i}\left( 2a_{i}\rho a_{i}^{\dag }-a_{i}^{\dag }a_{i}\rho -\rho a_{i}^{\dag }a_{i}\right) $, where $\rho$ is the density matrix of the system.
For simplicity, here, we set $\kappa_1=\kappa_2=\kappa$ (the discussions on $\kappa_1\neq \kappa_2$ are given in the Appendix~\ref{AppB}),
and rescale other parameters by $\kappa$, such as $\varepsilon=\kappa/10$ for weak pumping.

\begin{figure}[tbp]
\includegraphics[bb=40 198 545 621, width=8.5 cm, clip]{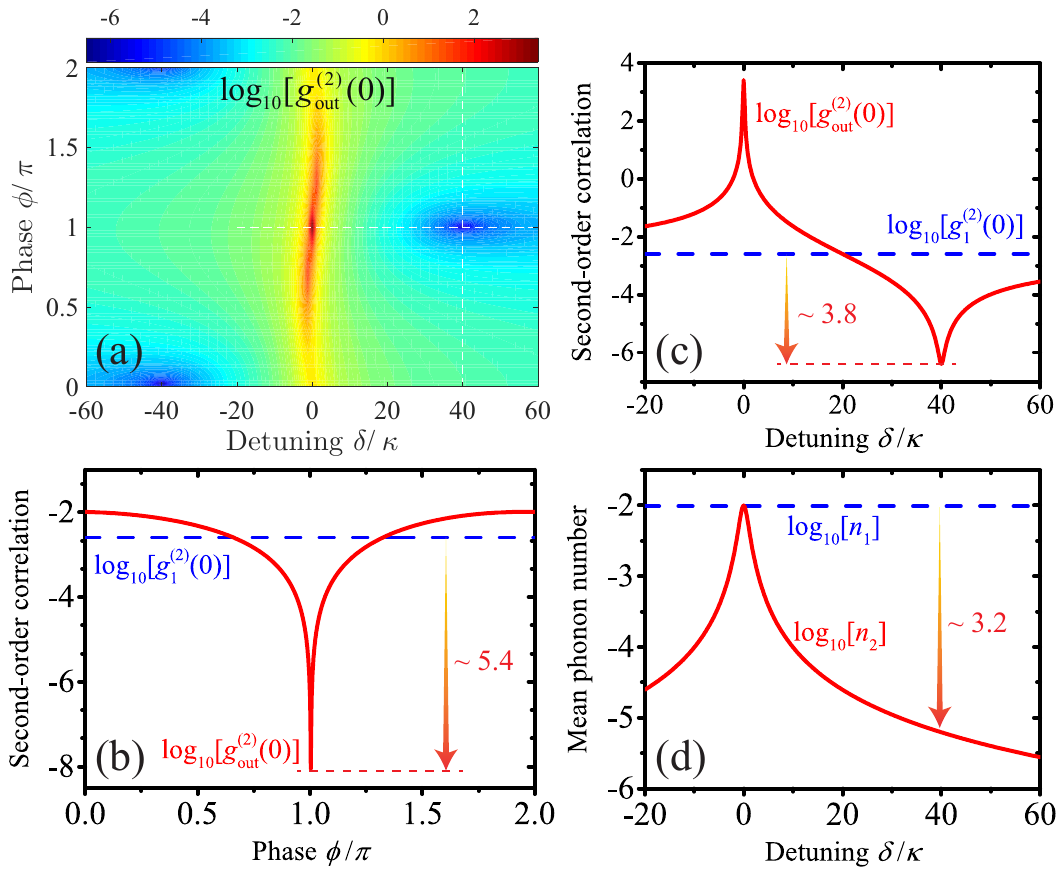}
\caption{(Color online) (a) The second-order correlation $\log_{10}[g^{(2)}_{\rm out}(0)]$ for different phase $\phi/\pi$ and detuning $\delta/\kappa$. The second-order correlations $\log_{10}[g^{(2)}_{\rm out}(0)]$ and $\log_{10}[g^{(2)}_{1}(0)]$ (b) versus phase $\phi/\pi$ with $\delta=2U$ and (c) versus detuning $\delta/\kappa$ with $\phi=\pi$. (d) The mean photon number $\log_{10}(n_{1})$ and $\log_{10}(n_{2})$ versus detuning $\delta/\kappa$ with $\phi=\pi$. The other parameters are $\Delta_1=0$, $U=20\kappa$, and $\varepsilon=0.1\kappa$.}
\label{fig2}
\end{figure}

\section{Photon Blockade Enhancement}

To demonstrate the photon blockade enhancement more clearly, the second-order correlation functions for photons in the two cavities $g_{i}^{( 2) }( 0)$ and in the output field $g_{\rm out}^{( 2) }( 0)$ are shown in Fig.~\ref{fig1}(b). As expected, the photon statistics in the two uncoupled cavities are independent of each other: strong photon blockade $g^{(2)}_1(0)\approx 2.54 \times 10^{-3}$ in the cavity $a_1$ for strong enharmonicity ($U=20\kappa$), and no photon blockade $g^{(2)}_2(0)=1$ in the cavity $a_2$ without nonlinearity. Surprisingly, a much stronger photon blockade is obtained in the mixing output field $a_{\mathrm{out}}$ for $[g_{\rm out}^{( 2) }( 0) / g^{(2)}_1(0)]\sim 10^ {-4}$ at resonant frequency ($\Delta_1=0$).
Moreover, as there is no direct coupling between the two cavities ($a_1$ and $a_2$), there is no oscillation in $g_{\rm out}^{( 2) }(\tau)$ [Fig.~\ref{fig1}(c)], in contrast to the rapid oscillations in photon correlations as a result of amplitude oscillation between different cavities~\cite{Bamba2011PRA}.

Figure~\ref{fig2}(a) is a color plot of $\log_{10}[g^{(2)}_{\rm out}(0)]$ as a function of the phase $\phi/\pi$ and detuning $\delta/\kappa$, for $\Delta_1=0$ and $U=20\kappa$.
The minimum of $\log_{10}[g^{(2)}_{\rm out}(0)]$ is reached for $\phi\approx\pi$ and $\delta\approx 2U$ (or $\phi\approx 0$ and $\delta\approx -2U$).
Two cuts taken from the color plot for $\delta=2U$ and $\phi=\pi$ are shown in Figs.~\ref{fig2}(b) and \ref{fig2}(c), respectively.
The photon blockade is enhanced, i.e. $g^{(2)}_{\rm out}(0)<g^{(2)}_{1}(0)$, in the regime of $0.66\pi <\phi<1.33 \pi$, and the minimal value of $g^{(2)}_{\rm out}(0)$ is about $5.4$ orders smaller than $g^{(2)}_{1}(0)$ at $\phi\approx 0.996\pi$ [Fig.~\ref{fig2}(b)].
Moreover, $g^{(2)}_{\rm out}(0)$ also strongly depends on the detuning $\delta/\kappa$ between the two cavities, and it is about $3.8$ orders smaller than $g^{(2)}_{1}(0)$ at $\delta\approx2U$ [Fig.~\ref{fig2}(c)].
These results suggested that quantum interference might be responsible for the great enhancement of photon blockade in the output field.

In addition, the mean photon numbers in the cavities ($n_i=\langle a_{i}^{\dag}a_{i} \rangle$) are plotted as functions of the detuning $\delta/\kappa$ in Fig.~\ref{fig2}(d).
Under the optimal condition $\delta\approx 2U$, the single photon generation in the auxiliary cavity $a_{2}$ is suppressed seriously ($n_2/n_1\approx10^{-3.2}$) for large detuning $\delta \gg \kappa$. 
Hence, almost all of the single photons in the mixing output field are emitted from the cavity $a_{1}$, and they are about one half of the photons emitted from cavity $a_{1}$.
Nevertheless, the auxiliary cavity $a_{2}$ provides another path for two photons passing through the whole system, which is the key ingredient for the enhancement of photon blockade in the output field as discussed bellow.
By the way, the mean photon number in the auxiliary cavity $a_{2}$ is almost the same as the one in cavity $a_{1}$ ($n_1\approx n_2$) under the resonant condition $\delta=0$, and they cancel each other at $\phi=\pi$ for destructive interference, which induces a strong bunching effect [$g^{(2)}_{\rm out}(0)\gg 1$]~\cite{Xu2014PRA2,Zubizarreta2020LPRv} in the output field [Fig.~\ref{fig2}(c)].

\section{Biquadratic scaling}

In order to understand the origin of the giant enhancement of photon blockade in the output field, we derive the expressions of the second-order correlations [$g^{(2)}_{\rm out}(0)$ and $g^{(2)}_{1}(0)$] analytically (see Appendix~\ref{AppB}).
Here, including the effect of optical decay, an effective Hamiltonian $H_{\rm eff}=H-i\kappa(a_{1}^{\dag}a_{1}+a_{2}^{\dag}a_{2})$ is introduced according to the quantum-trajectory method~\cite{Plenio1998RMP}. Under weak driving condition ($\varepsilon \ll \kappa$), the wave function on a Fock-state basis can be truncated to the two-photon manifold as: $\left\vert \psi \right\rangle
=\sum_{n_{2}=0}^{2-n_{1}}\sum_{n_{1}=0}^{2}C_{n_{1}n_{2}}\left\vert
n_{1},n_{2}\right\rangle $. Here, $\left\vert n_{1},n_{2}\right\rangle$ represents the Fock state of $n_1$ photons in cavity $a_1$ and  $n_2$ photons in cavity $a_2$, with the probability amplitude $C_{n_{1}n_{2}}$.
The expression of $C_{n_{1}n_{2}}$ can be obtained by solving the Shr\"{o}dinger equation $d\left\vert \psi \right\rangle /dt=-iH_{\mathrm{eff}}\left\vert \psi \right\rangle $ in the steady states. 

\begin{figure}[tbp]
\centering
\includegraphics[bb=175 337 407 528, width=7cm, clip]{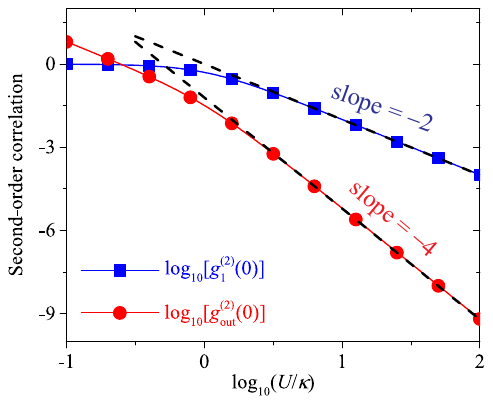}
\caption{(Color online) The second-order correlations $\log_{10}[g^{(2)}_{\rm out}(0)]$ and $\log_{10}[g^{(2)}_{1}(0)]$ versus the nonlinear interaction strength $\log_{10}(U/\kappa)$ with $\Delta_1=0$, $\phi=\pi$, and $\delta=2U$.}
\label{fig3}
\end{figure}

Under the conditions for photon blockade enhancement ($\phi=\pi$, $\Delta_1=0$, and $\delta=2U\gg \kappa$), the second-order correlation function can be written as
\begin{eqnarray}\label{Eq4}
g_{\mathrm{out}}^{\left( 2\right) }\left( 0\right)\approx \frac{2\kappa
^{2}}{N_{\mathrm{out}}^{2}}  && \left\{ \left\vert C_{20}-\sqrt{2}%
C_{11}\right\vert ^{2}+\left\vert C_{02}\right\vert ^{2}\right.  \nonumber \\
&&\left. -2{\rm Re}\left[ \left( \sqrt{2}C_{11}^{\ast }-C_{20}^{\ast
}\right) C_{02}\right] \right\},
\end{eqnarray}%
where $N_{\mathrm{out}}\approx \kappa [ \left\vert C_{10}\right\vert
^{2}+\left\vert C_{01}\right\vert ^{2}-2{\rm Re}\left( C_{01}C_{10}^{\ast
}\right)]$. The probability amplitudes for two-photon states in the steady state are approximately given by
\begin{eqnarray}
C_{20} &\approx &\frac{-i}{U-i\kappa }\frac{\varepsilon ^{2}}{\sqrt{2}\kappa 
},\qquad C_{02}\approx -\frac{\sqrt{2}}{2U-i\kappa }\frac{\varepsilon ^{2}}{%
4U}, \nonumber\\
C_{11} &\approx &\frac{-i}{U-i\kappa }\left( \frac{\varepsilon ^{2}}{2\kappa 
}-\frac{i\varepsilon ^{2}}{4U}\right)
\end{eqnarray}
with $|C_{20}|\approx \sqrt{2} |C_{11}| \gg |C_{02}|$.
The first and last terms inside the curly brace of Eq.~(\ref{Eq4}) are canceled out by the destructive interference between $C_{20}$ and $C_{11}$.
Then the second-order correlation function in the output field is approximately given by
\begin{equation}\label{Eq5}
g_{\mathrm{out}}^{\left( 2\right) }\left( 0\right) \approx [\kappa /(2U)] ^{4}.
\end{equation}
The second-order correlation of the photons in the output field depends on the strength of the nonlinear interaction with a biquadratic scaling law, which is different from the second-order correlation of photons in the cavity $a_1$, i.e., $g_{\mathrm{1}}^{( 2)}( 0) \approx ( \kappa/U) ^{2}$, with a quadratic scaling law.

Both $\log_{10}[g^{(2)}_{\rm out}(0)]$ and $\log_{10}[g^{(2)}_{1}(0)]$, obtained by solving the master equation numerically, are plotted as functions of $\log_{10}(U/\kappa)$ in Fig.~\ref{fig3}. In the strong nonlinear regime $U/\kappa \gg 1$, the slope of $\log_{10}[g^{(2)}_{\rm out}(0)]$ versus $\log_{10}(U/\kappa)$ is $-4$, which is much larger than the slope of $-2$ for $\log_{10}[g^{(2)}_{1}(0)]$ versus $\log_{10}(U/\kappa)$. The numerical results agree well with the analytical expressions in the strong nonlinear regime (black dashed lines in Fig.~\ref{fig3}).
Thus, the scheme we proposed can not only greatly enhance photon blockade by several orders, but also change the scaling exponent of the second-order correlation on the nonlinear interaction strength from $-2$ to $-4$.


\section{JC model}

The scheme for photon blockade enhancement in output field is general. It can be extended to other optical platforms with anharmonic energy levels, such as the JC model~\cite{JCModel}. The strong coupling between a single cavity and a TLS has been realized decades ago~\cite{Thompson1992PRL,Wallraff2004Natur,Reithmaier2004Natur,Yoshie2004Natur}, and photon blockade was demonstrated in a large number of experiments based on JC model~\cite{Birnbaum2005Natur,Dayan2008Sci,Aoki2009PRL,Faraon2008NatPh,Reinhard2012NaPho,LangC2011PRL,Hoffman2011PRL}. 
Here, we demonstrate a scaling enhancement of photon blockade in the mixing field output from two (uncoupled) cavities with a TLS strongly coupled to one of them.

\begin{figure}[tbp]
\includegraphics[bb=36 187 552 808, width=8.5 cm, clip]{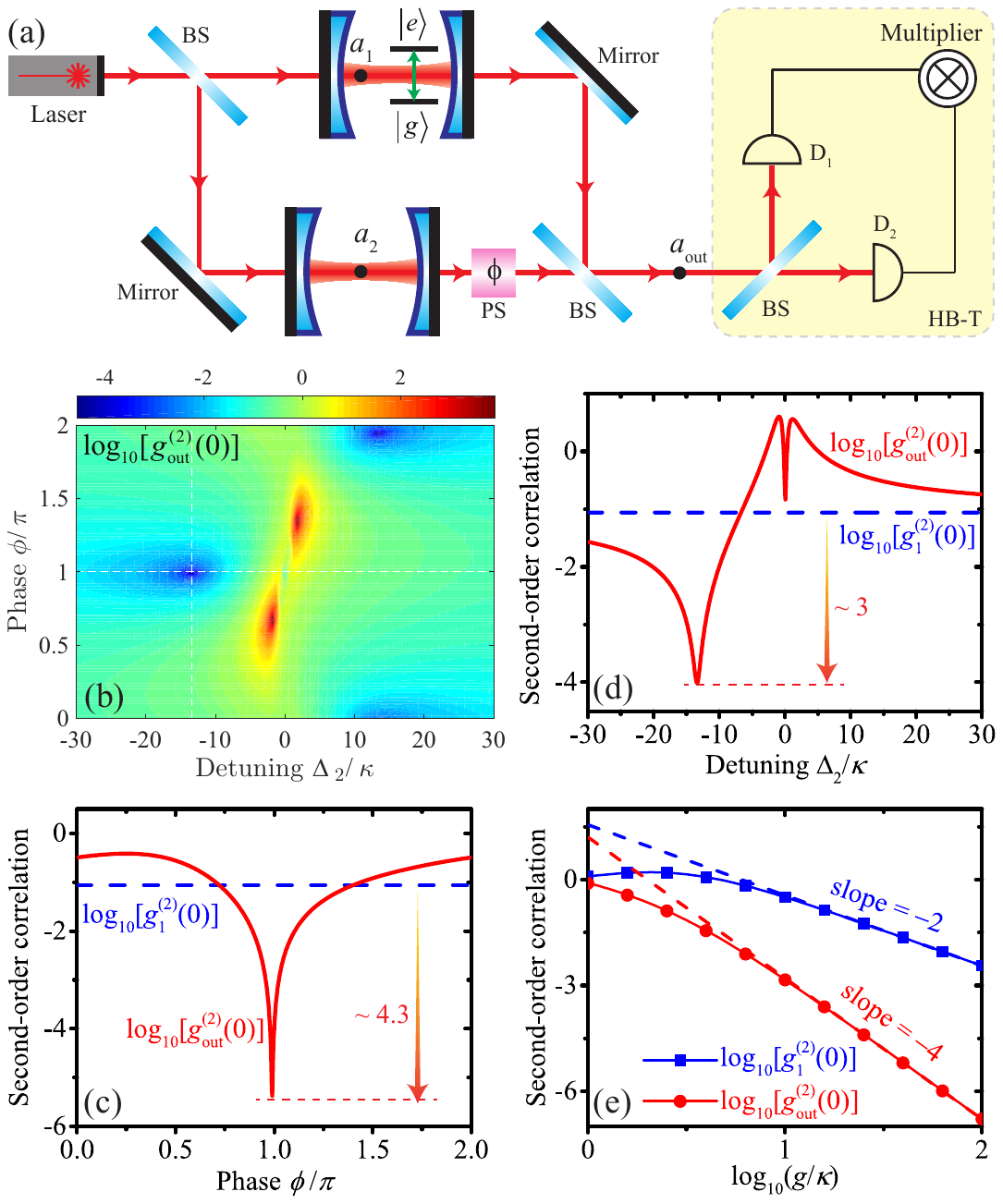}
\caption{(Color online) Photon blockade enhancement in the mixing output field of two cavities with a TLS in one of them.
(a) Sketch of the proposed scheme with a TLS (excited state $|e\rangle$ and ground state $|g\rangle$) in cavity $a_1$.
(b) The second-order correlation $\log_{10}[g^{(2)}_{\rm out}(0)]$ for different phase $\phi/\pi$ and detuning $\Delta_2/\kappa$. The second-order correlations $\log_{10}[g^{(2)}_{\rm out}(0)]$ and $\log_{10}[g^{(2)}_{1}(0)]$ (c) versus phase $\phi/\pi$ with $\Delta_2=-2g/3$ and (d) versus detuning $\Delta_2/\kappa$ with $\phi=\pi$. (e) The second-order correlations $\log_{10}[g^{(2)}_{\rm out}(0)]$ and $\log_{10}[g^{(2)}_{1}(0)]$ versus the interaction strength $\log_{10}(g/\kappa)$ with $\phi=\pi$, $\Delta_1=-g$ and $\Delta_2=-2g/3$. The other parameters are $g=20\kappa$ and $\kappa_a=2\kappa$.
}
\label{fig4}
\end{figure}

The scheme can be extended to the JC model just by replacing the $\chi^{(3)}$ nonlinear medium [Fig.~\ref{fig1}(a)] by a TLS [Fig.~\ref{fig4}(a)], and the system is described by
\begin{eqnarray}
H_{\mathrm{JC}} &=&\Delta_1 a_{1}^{\dag }a_{1}+\Delta _{a}\sigma
_{+}\sigma _{-}+g\left( a_{1}^{\dag }\sigma _{-}+\sigma _{+}a_{1}\right)  \nonumber\\
&&+\Delta_2 a_{2}^{\dag }a_{2}+i\varepsilon\left( a_{1}^{\dag
}+ a_{2}^{\dag }-{\rm H.c.}\right) ,
\end{eqnarray}
where $\sigma _{+}$ and $\sigma _{-}$ are the raising and lowering operators of the TLS with transition frequency $\omega_a$, $\Delta_a=\omega_a - \omega_p$ is the laser detuning from the TLS, and $g$ is the TLS-cavity coupling strength.
We assume that the TLS is resonant with the cavity ($\Delta_a=\Delta_1=\Delta$), and the decay rate of the TLS is $\kappa_a=2\kappa$.

In order to confirm the applicability of our scheme in JC model, we perform a fully numerical simulation of the second-order correlation in the output field based on the master equation. $\log_{10}[g^{(2)}_{\rm out}(0)]$ is plotted as a function of $\phi/\pi$ and $\Delta_2/\kappa$ in Fig.~\ref{fig4}(b), for $g=20\kappa$ and $\Delta=-g$. The minimum of $\log_{10}[g^{(2)}_{\rm out}(0)]$ appears around $\phi=\pi$ and $\Delta_2\approx -13.3\kappa \approx -2g/3$. From the cuts of the color scale plot shown in Figs.~\ref{fig4}(c) and \ref{fig4}(d), $g^{(2)}_{\rm out}(0)$ is about $4.3$ orders smaller than $g^{(2)}_{1}(0)$ at $\phi=0.99\pi$ with $\Delta_2=-2g/3$ [Fig.~\ref{fig4}(c)], and about 3 orders smaller than $g^{(2)}_{1}(0)$ at $\Delta_2 \approx -13.38\kappa$ with $\phi=\pi$ [Fig.~\ref{fig4}(d)].

In order to understand the enhancement of photon blockade in the output field, we derive the expression of $g^{(2)}_{\rm out}(0)$ by using the effective Hamiltonian $H_{\rm JC,eff}=H_{\rm JC}-i\kappa(a_{1}^{\dag}a_{1}+a_{2}^{\dag}a_{2}+\sigma _{+}\sigma
_{-})$ and wave function $\left\vert \varphi \right\rangle
=\sum_{n_{2}=0}^{2-n_{1}}\sum_{n_{1}=0}^{2}C_{gn_{1}n_{2}}\left\vert
g,n_{1},n_{2}\right\rangle +\sum_{n_{2}=0}^{1-n_{1}}\sum_{n_{1}=0}^{1}C_{en_{1}n_{2}}\left\vert
e,n_{1},n_{2}\right\rangle$ (see Appendix~\ref{AppC}).
Here, $\left\vert g, n_{1},n_{2}\right\rangle$ ($\left\vert e, n_{1},n_{2}\right\rangle$) denotes the Fock state of $n_1$ photons in cavity $a_1$, $n_2$ photons in cavity $a_2$, and the TLS in the ground (excited) state, with the probability amplitude $C_{gn_{1}n_{2}}$ ($C_{en_{1}n_{2}}$).
The optimal condition $\Delta_2=-2g/3$ for photon blockade in the output field is obtained by setting $C_{g02}\approx \sqrt{2}C_{g11}$, and they are canceled out by the destructive interference in the output field at phase difference $\phi=\pi$.
Thus the second-order correlation function in the output field becomes (see more details in Appendix~\ref{AppC}):
\begin{equation}
g_{\mathrm{out}}^{\left( 2\right) }\left( 0\right) \approx 16(\kappa/g) ^{4}.
\end{equation}
We also have the second-order correlation in cavity $a_1$ as $g_{\mathrm{1}}^{( 2) }( 0) \approx 36( \kappa/g)^{2}$ for single-mode JC model, and
they [dash lines in Fig.~\ref{fig4}(e)] agree well with numerical results in the strong coupling regime.
Similar to the case for the cavity containing $\chi^{(3)}$ nonlinear medium, the scheme with JC model can also enhance photon blockade in the output field by several orders, and change the scaling exponent of the second-order correlation on the strength of the TLS-cavity interaction from $-2$ to $-4$.

\section{Conclusions}

In conclusion, we have proposed a scheme to achieve scaling enhancement of photon blockade in the mixing field output from a nonlinear cavity (in the strong nonlinear regime) and an auxiliary (linear) cavity.
We identify that the probability for two photons in the output field can be significantly inhibited by the quantum interference between two of the paths for two photons passing through the whole system, leading to a biquadratic scaling relation between the second-order correlation of the photons in the output field and intracavity nonlinear interaction strength, in contrast to a quadratic scaling relation for the photons in a nonlinear cavity.
The scheme for photon blockade enhancement is general, for it not only achievable in the cavity containing $\chi^{(3)}$ nonlinearity~\cite{ChenJH2019PRL} and TLS~\cite{Walther_2006,GU20171},
but also applicable in cavities with other nonlinear interactions, such as $\chi^{(2)}$ nonlinearity~\cite{Zhang2019NaPho,Lu2020OPTICA} and optomechanical interactions~\cite{Aspelmeyer2014RMP,Rabl2011PRL,Nunnenkamp2011PRL,Liao2013PRA,Kronwald2013PRA,XuXW2013PRA,Das2023PRL} (see Appendixes~\ref{AppD} and \ref{AppE}).
Furthermore, our scheme can be directly extended to enhance phonon blockade~\cite{LiuYX2010PRA,XuXW2016PRA,Debnath2018PRL,Patil2022PRL}, magnon blockade~\cite{Zhang2016SciA,LiuZX2019PRB,Yuan2020PRB,XieJK2020PRA,JinZ2023PRA}, and polariton blockade~\cite{Verger2006PRB,HuaiSN2018PRA,Denning2011PRR,munoz2019NM,delteil2019NM}.
It is worth mentioning that, as the second-order correlations become very small, there are some other noises in the experiments, such as the noises in the lasers and photodetectors, that may weaken the photon blockade effect, and such effect should be considered case by case.



\begin{acknowledgments}
This work is supported by the National Natural Science Foundation of China (Grants No.~12064010, No.~12247105, and No.~12421005),
the science and technology innovation Program of Hunan Province (Grant No.~2022RC1203), 
and Hunan provincial major sci-tech program (Grant No.~2023ZJ1010).
\end{acknowledgments}

\appendix

\section{Second-order correlation of $A_{\mathrm{out}}$}

\label{AppA} 

\begin{figure*}[tbp]
\centering
\includegraphics[bb=9 388 575 554, width=16 cm, clip]{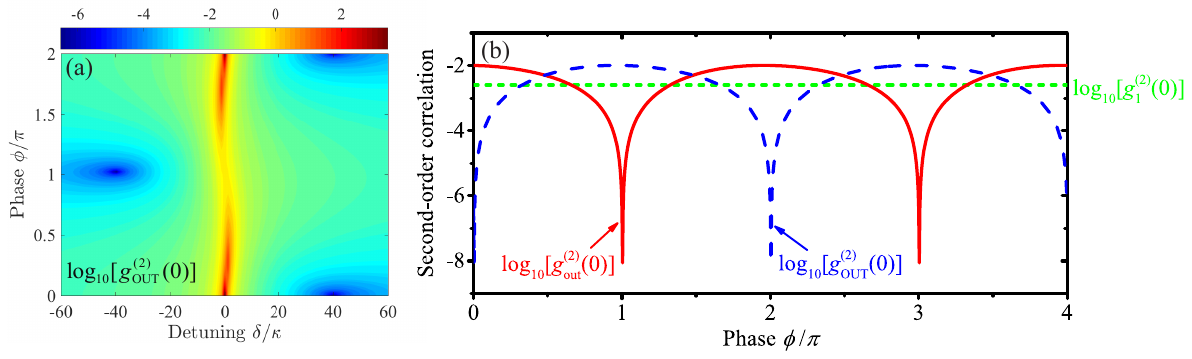}
\caption{(Color online) (a) The second-order correlation $\log_{10}[g^{(2)}_{%
\mathrm{OUT}}(0)]$ for different phase $\protect\phi/\protect\pi$ and
detuning $\protect\delta/\protect\kappa$. (b) The second-order correlations $%
\log_{10}[g^{(2)}_{\mathrm{out}}(0)]$, $\log_{10}[g^{(2)}_{\mathrm{OUT}}(0)]$%
, and $\log_{10}[g^{(2)}_{1}(0)]$ versus phase $\protect\phi/\protect\pi$
with $\protect\delta=2U$. The other parameters are $\Delta_1=0$, $U=20%
\protect\kappa$, and $\protect\varepsilon=0.1\protect\kappa$.}
\label{fig5}
\end{figure*}

The second-order correlation function of the output field $A_{%
\mathrm{out}}$ is defined by%
\begin{eqnarray}  \label{g_OUT}
g_{\mathrm{OUT}}^{\left( 2\right) }\left( \tau \right) &\equiv&\frac{%
\left\langle A_{\mathrm{out}}^{\dag }A_{\mathrm{out}}^{\dag }(\tau )A_{%
\mathrm{out}}(\tau )A_{\mathrm{out}}\right\rangle }{\left\langle A_{\mathrm{%
out}}^{\dag }A_{\mathrm{out}}\right\rangle ^{2}} \\
&=&\sum_{j,k,l,m=1}^{2}e^{in\phi^{\prime}}\sqrt{\kappa _{j}\kappa _{k}\kappa
_{l}\kappa _{m}}\frac{\left\langle a_{j}^{\dag }a_{k}^{\dag }(\tau
)a_{l}(\tau )a_{m}\right\rangle }{N_{\mathrm{OUT}}^{2}},  \notag
\end{eqnarray}%
where $n=l+m-j-k$, $\phi^{\prime}= \phi +\pi$, and $N_{\mathrm{OUT}}=\kappa
_{1}\langle a_{1}^{\dag }a_{1}\rangle +\kappa _{2}\langle a_{2}^{\dag
}a_{2}\rangle +2\sqrt{\kappa _{1}\kappa _{2}}\mathrm{Re}( e^{i( \phi +\pi )
}\langle a_{1}^{\dag }a_{2}\rangle ) $. The second-order correlation $%
g^{(2)}_{\mathrm{OUT}}(0)$ can be obtained from $g^{(2)}_{\mathrm{out}}(0)$
just with $\phi $ replaced by $\phi^{\prime}$.

The second-order correlation $\log_{10}[g^{(2)}_{\mathrm{OUT}}(0)]$ for
different phase $\phi/\pi$ and detuning $\delta/\kappa$ are shown in Fig.~%
\ref{fig5}(a). Different from the second-order correlation $%
\log_{10}[g^{(2)}_{\mathrm{out}}(0)]$ [Fig.~\ref{fig2}(a)], the
minimal values of $\log_{10}[g^{(2)}_{\mathrm{OUT}}(0)]$ are obtained for $%
\{\delta\approx -2U,\; \phi\approx \pi\}$ or $\{\delta\approx 2U,\;
\phi\approx 0\}$. In order to show the relation between $\log_{10}[g^{(2)}_{%
\mathrm{out}}(0)]$ and $\log_{10}[g^{(2)}_{\mathrm{OUT}}(0)]$ clearly, we
plot $\log_{10}[g^{(2)}_{\mathrm{out}}(0)]$, $\log_{10}[g^{(2)}_{\mathrm{OUT}%
}(0)]$, and $\log_{10}[g^{(2)}_{1}(0)]$ versus phase $\phi/\pi$ in Fig.~\ref%
{fig5}(b). In comparing with $\log_{10}[g^{(2)}_{1}(0)]$, $%
\log_{10}[g^{(2)}_{\mathrm{out}}(0)]$ and $\log_{10}[g^{(2)}_{\mathrm{OUT}%
}(0)]$ are enhanced or suppressed periodically with the phase $\phi/\pi$,
and there is a phase difference of $\pi$ between them, which are consistent
with Eqs.~(\ref{g_out}) and (\ref{g_OUT}). To avoid unnecessary duplication,
we focus on the output field $a_{\mathrm{out}}$ in the main text.

\section{$\protect\chi^{(3)}$ nonlinearity}

\label{AppB}

In this Appendix, we will derive the analytical expressions of the
second-order correlation function for photons in the mixing output field of
two (uncoupled) cavities with $\chi ^{(3)}$ nonlinearity in one of them. In
order to obtain the analytical expression of the second-order correlation
function of the photons in the output field, we use the wave function:%
\begin{eqnarray}
\left\vert \psi \right\rangle &=&C_{00}\left\vert 0,0\right\rangle
+C_{10}\left\vert 1,0\right\rangle +C_{01}\left\vert 0,1\right\rangle  \notag
\\
&&+C_{20}\left\vert 2,0\right\rangle +C_{11}\left\vert 1,1\right\rangle
+C_{02}\left\vert 0,2\right\rangle +\cdots ,
\end{eqnarray}%
where $\left\vert n_{1},n_{2}\right\rangle $ represents the Fock state with $%
n_{1}$ photons in cavity $a_{1}$ and $n_{2}$ photons in cavity $a_{2}$, with
the probability amplitude $C_{n_{1}n_{2}}$. According to the
quantum-trajectory method~\cite{Plenio1998RMP}, we introduce an effective
Hamiltonian 
\begin{eqnarray}
H_{\mathrm{eff}} &=&H-i\kappa _{1}a_{1}^{\dag }a_{1}-i\kappa _{2}a_{2}^{\dag
}a_{2}  \notag  \label{3thr} \\
&=&(\Delta _{1}-i\kappa _{1})a_{1}^{\dag }a_{1}+Ua_{1}^{\dag }a_{1}^{\dag
}a_{1}a_{1}+i\varepsilon _{1}(a_{1}^{\dag }-a_{1})  \notag \\
&&+(\Delta _{2}-i\kappa _{2})a_{2}^{\dag }a_{2}+i\varepsilon
_{2}(a_{2}^{\dag }-a_{2}),
\end{eqnarray}%
to include the decay effect of cavities. By substituting the wave function
and effective Hamiltonian into the Schr\"{o}dinger equation $id\left\vert
\psi \right\rangle /dt=H_{\mathrm{eff}}\left\vert \psi \right\rangle $, we
get the dynamic equations for the probability amplitude $C_{n_{1}n_{2}}$ as 
\begin{eqnarray}
i\frac{d}{dt}C_{10} &=&\left( \Delta _{1}-i\kappa _{1}\right)
C_{10}+i\varepsilon _{1}C_{00}-i\varepsilon _{2}C_{11}-i\sqrt{2}\varepsilon
_{1}C_{20},  \notag \\
i\frac{d}{dt}C_{01} &=&\left( \Delta _{2}-i\kappa _{2}\right)
C_{01}+i\varepsilon _{2}C_{00}-i\varepsilon _{1}C_{11}-i\sqrt{2}\varepsilon
_{2}C_{02},  \notag \\
i\frac{d}{dt}C_{20} &=&\left( 2\Delta _{1}+2U-i2\kappa _{1}\right) C_{20}+i%
\sqrt{2}\varepsilon _{1}C_{10},  \notag \\
i\frac{d}{dt}C_{02} &=&\left( 2\Delta _{2}-i2\kappa _{2}\right) C_{02}+i%
\sqrt{2}\varepsilon _{2}C_{01},  \notag \\
i\frac{d}{dt}C_{11} &=&\left( \Delta _{1}+\Delta _{2}-i\kappa _{1}-i\kappa
_{2}\right) C_{11}+i\varepsilon _{2}C_{10}+i\varepsilon _{1}C_{01},  \notag
\\
\cdots &&
\end{eqnarray}%
In the steady state, e.g., $dC_{n_{1}n_{2}}/dt=0$, we have 
\begin{eqnarray}
-i\varepsilon _{1}C_{00} &=&\left( \Delta _{1}-i\kappa _{1}\right)
C_{10}-i\varepsilon _{2}C_{11}-i\sqrt{2}\varepsilon _{1}C_{20},  \notag \\
-i\varepsilon _{2}C_{00} &=&\left( \Delta _{2}-i\kappa _{2}\right)
C_{01}-i\varepsilon _{1}C_{11}-i\sqrt{2}\varepsilon _{2}C_{02},  \notag \\
0 &=&\left( 2\Delta _{1}+2U-i2\kappa _{1}\right) C_{20}+i\sqrt{2}\varepsilon
_{1}C_{10},  \notag \\
0 &=&\left( 2\Delta _{2}-i2\kappa _{2}\right) C_{02}+i\sqrt{2}\varepsilon
_{2}C_{01},  \notag \\
0 &=&\left( \Delta _{1}+\Delta _{2}-i\kappa _{1}-i\kappa _{2}\right)
C_{11}+i\varepsilon _{2}C_{10}+i\varepsilon _{1}C_{01},  \notag \\
&&\cdots
\end{eqnarray}%
Under weak driving conditions $\left\{ \varepsilon _{1},\varepsilon
_{2}\right\} \ll \{\kappa _{1},\kappa _{2}\}$, we have $|C_{00}|\approx 1\gg
\{|C_{10}|,|C_{01}|\}\gg \{|C_{20}|,|C_{02}|,|C_{11}|\}$, then the
probability amplitudes of one-photon states ($C_{10}$ and $C_{01}$) are
obtained as 
\begin{equation}
C_{10}=\frac{-i\varepsilon _{1}}{\Delta _{1}-i\kappa _{1}},
\end{equation}%
\begin{equation}
C_{01}=\frac{-i\varepsilon _{2}}{\Delta _{2}-i\kappa _{2}},
\end{equation}%
and the probability amplitudes of two-photon states ($C_{20}$, $C_{11}$, and 
$C_{02}$) are obtained as 
\begin{equation}  \label{S15}
C_{20}=\frac{-i\sqrt{2}\varepsilon _{1}}{2\Delta _{1}+2U-i2\kappa _{1}}%
C_{10},
\end{equation}%
\begin{equation}
C_{11}=\frac{-i\varepsilon _{2}C_{10}-i\varepsilon _{1}C_{01}}{\Delta
_{1}+\Delta _{2}-i\left( \kappa _{1}+\kappa _{2}\right) },
\end{equation}%
\begin{equation}  \label{S17}
C_{02}=\frac{-i\sqrt{2}\varepsilon _{2}}{2\Delta _{2}-i2\kappa _{2}}C_{01}.
\end{equation}

Under weak driving conditions, the first-order correlation functions can be
given by the steady state probability amplitudes $C_{n_{1}n_{2}}$ as 
\begin{eqnarray}
\left\langle a_{1}^{\dag }a_{1}\right\rangle  &\approx &\left\vert
C_{10}\right\vert ^{2},\quad \left\langle a_{2}^{\dag }a_{2}\right\rangle
\approx \left\vert C_{01}\right\vert ^{2},  \notag \\
\quad \left\langle a_{1}^{\dag }a_{2}\right\rangle  &\approx
&C_{01}C_{10}^{\ast },
\end{eqnarray}%
and the second-order correlation functions can be given by 
\begin{eqnarray}
\left\langle a_{1}^{\dag }a_{1}^{\dag }a_{1}a_{1}\right\rangle  &\approx
&2\left\vert C_{20}\right\vert ^{2},\left\langle a_{1}^{\dag }a_{1}^{\dag
}a_{1}a_{2}\right\rangle \approx \sqrt{2}C_{11}C_{20}^{\ast },  \notag \\
\left\langle a_{2}^{\dag }a_{2}a_{1}^{\dag }a_{1}\right\rangle  &\approx
&\left\vert C_{11}\right\vert ^{2},\left\langle a_{1}^{\dag }a_{1}^{\dag
}a_{2}a_{2}\right\rangle \approx 2C_{02}C_{20}^{\ast }, \\
\left\langle a_{2}^{\dag }a_{1}^{\dag }a_{2}a_{2}\right\rangle  &\approx &%
\sqrt{2}C_{02}C_{11}^{\ast },\left\langle a_{2}^{\dag }a_{2}^{\dag
}a_{2}a_{2}\right\rangle \approx 2\left\vert C_{02}\right\vert ^{2},  \notag
\end{eqnarray}%
Thus, the second-order correlation function in the output field can be given
approximately as 
\begin{eqnarray} \label{g2}
g_{\mathrm{out}}^{\left( 2\right) }\left( 0\right)  &\approx &\frac{2}{N_{R,%
\mathrm{out}}^{2}}\left\{ \left\vert \kappa _{1}C_{20}+e^{i\phi }\sqrt{%
2\kappa _{1}\kappa _{2}}C_{11}\right\vert ^{2}+\left\vert \kappa
_{2}C_{02}\right\vert ^{2}\right.   \notag \\
&&\left. +2\mathrm{Re}\left[ \left( \kappa _{1}C_{20}+e^{i\phi }\sqrt{%
2\kappa _{1}\kappa _{2}}C_{11}\right) e^{-i2\phi }\kappa _{2}C_{02}^{\ast }%
\right] \right\} ,\notag \\
\end{eqnarray}%
where $N_{\mathrm{out}}\approx \left\vert \sqrt{\kappa _{1}}C_{10}+e^{i\phi }%
\sqrt{\kappa _{2}}C_{01}\right\vert ^{2}$.

First of all, let us consider the conditions that $\varepsilon
_{1}=\varepsilon _{2}=\varepsilon $ and $\kappa _{1}=\kappa _{2}=\kappa $.
According to the numerical results shown in Fig.~2 in the main text, photon
blockade in the output field is enhanced greatly under the conditions $%
\{\delta =2U,\;\phi =\pi \}$ or $\{\delta =-2U,\;\phi =0\}$, with the other
parameters $\{\Delta _{1}=0$ and $U\gg \kappa \}$. In order to understand
these phenomena analytically, we rewrite the second-order correlation as%
\begin{eqnarray}
g_{\mathrm{out}}^{\left( 2\right) }\left( 0\right)  &\approx &\frac{2\kappa
^{2}}{N_{\mathrm{out}}^{2}}\left\{ \left\vert C_{20}+e^{i\phi }\sqrt{2}%
C_{11}\right\vert ^{2}+\left\vert C_{02}\right\vert ^{2}\right.   \\
&&\left. +2\mathrm{Re}\left[ \left( C_{20}+e^{i\phi }\sqrt{2}C_{11}\right)
\left( e^{i2\phi }C_{02}\right) ^{\ast }\right] \right\} \notag 
\end{eqnarray}%
where%
\begin{equation}
C_{10}=\frac{\varepsilon }{\kappa },\qquad C_{01}\approx -i\frac{\varepsilon 
}{2U},
\end{equation}%
and%
\begin{eqnarray}
C_{20} &\approx &\frac{-i\sqrt{2}\varepsilon }{2U-i2\kappa }\frac{%
\varepsilon }{\kappa }, \\
C_{11} &\approx &\left( \frac{-\varepsilon ^{2}}{\delta -i2\kappa }\right)
\left( \frac{i}{\kappa }+\frac{1}{2U}\right) , \\
C_{02} &\approx &-\frac{\sqrt{2}\varepsilon }{4U-i2\kappa }\frac{\varepsilon 
}{2U}.
\end{eqnarray}%
In the strong nonlinear regime, i.e., $\left\vert \delta \right\vert =2U\gg
\kappa $, we have%
\begin{equation}
\left\vert C_{10}\right\vert \gg \left\vert C_{01}\right\vert ,\qquad
\left\vert C_{20}\right\vert \approx \left\vert \sqrt{2}C_{11}\right\vert
\gg \left\vert C_{02}\right\vert .
\end{equation}%
Under the conditions $\{\delta =2U,\;\phi =\pi \}$ or $\{\delta =-2U,\;\phi
=0\}$, $C_{20}$ and $\sqrt{2}C_{11}$ cancel each other out by destructive
interference, with 
\begin{equation}
\left\vert C_{20}+e^{i\phi }\sqrt{2}C_{11}\right\vert \ll \left\vert
C_{20}\right\vert \approx \left\vert \sqrt{2}C_{11}\right\vert .
\end{equation}%
In this case, the second-order correlation function for the photons in the
output field is given by 
\begin{equation}
g_{\mathrm{out}}^{\left( 2\right) }\left( 0\right) \approx \frac{1}{16}%
\left( \frac{\kappa }{U}\right) ^{4},
\end{equation}%
which is much smaller than the second-order correlation function for the
photons in the cavity $a_{1}$ 
\begin{equation}
g_{\mathrm{1}}^{\left( 2\right) }\left( 0\right) =\frac{\left\langle
a_{1}^{\dag }a_{1}^{\dag }a_{1}a_{1}\right\rangle }{\left\langle a_{1}^{\dag
}a_{1}\right\rangle ^{2}}\approx \frac{2\left\vert C_{20}\right\vert ^{2}}{%
\left\vert C_{10}\right\vert ^{4}}\approx \left( \frac{\kappa }{U}\right)
^{2}
\end{equation}%
in the strong nonlinear regime.

In addition, we discuss how to achieve scaling enhancement of photon
blockade with $\kappa _{1} \neq \kappa _{2}$. According to the definition of 
$\varepsilon _{i}$, i.e., $\varepsilon _{i}=\sqrt{\kappa _{i}P_{\mathrm{in}%
}/\hbar \omega_p}$, where $P_{\mathrm{in}}$ is the driving power of the two
cavities, we have $\varepsilon _{1}/\varepsilon _{2}=\sqrt{\kappa
_{1}/\kappa_{2}}$. According to Eq.~(\ref{g2}), in order to achieve scaling
enhancement of photon blockade, the coefficients ($C_{20}$, $C_{11}$, and $%
C_{02}$) should satisfy the conditions $\{|\kappa_{2}C_{02}|,|\kappa _{1}C_{20}+e^{i\phi }\sqrt{2\kappa _{1}\kappa
_{2}}C_{11}|\}\ll|\kappa _{1}C_{20}|\approx|\sqrt{2\kappa _{1}\kappa _{2}}%
C_{11}|$. Based on the expressions of the coefficients ($C_{20}$, $C_{11}$, and $%
C_{02} $) [Eqs.~(\ref{S15})-(\ref{S17})], the conditions are satisfied with 
\begin{equation}  \label{S30}
\Delta_1=0,\qquad \delta \approx 2U\kappa_2/\kappa_1,\qquad \phi \approx \pi,
\end{equation}
or 
\begin{equation}  \label{S31}
\Delta_1=0,\qquad \delta \approx -2U\kappa_2/\kappa_1,\qquad \phi \approx 0
\end{equation}
in the strong nonlinear regime ($U\gg\{\kappa_1,\kappa_2\}$) and the ratio $%
\kappa_2/\kappa_1\gg(\kappa_1/2U)^2$. In this case, the second-order
correlation function for the photons in the output field is given by 
\begin{equation}
g_{\mathrm{out}}^{\left( 2\right) }\left( 0\right) =\frac{1}{16}\left( \frac{%
\kappa _{1}}{U}\right) ^{4},
\end{equation}
which also depends on the strength of the nonlinear interaction $U$ with a
biquadratic scaling law.

\begin{figure*}[tbp]
\centering
\includegraphics[bb=19 467 555 589, width=18 cm, clip]{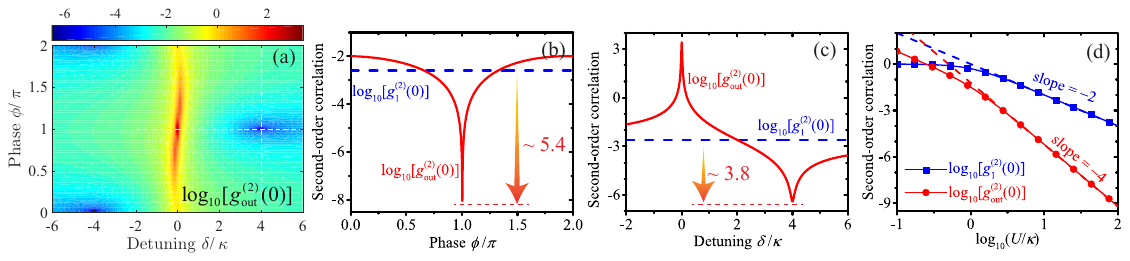}
\caption{(Color online) (a) The second-order correlation $\log_{10}[g^{(2)}_{%
\mathrm{out}}(0)]$ for different phase $\protect\phi/\protect\pi$ and
detuning $\protect\delta/\protect\kappa$ with $\protect\kappa_2=\protect%
\kappa_1/10$ and $U=20\protect\kappa$. The second-order correlations $%
\log_{10}[g^{(2)}_{\mathrm{out}}(0)]$ and $\log_{10}[g^{(2)}_{1}(0)]$ (b)
versus phase $\protect\phi/\protect\pi$ with $\protect\delta=2U\protect\kappa%
_2/\protect\kappa_1$ and $U=20\protect\kappa$, (c) versus detuning $\protect%
\delta/\protect\kappa$ with $\protect\phi=\protect\pi$ and $U=20\protect%
\kappa$, and (d) versus the $\protect\chi^{(3)}$ nonlinear interaction
strength $\log_{10}(U/\protect\kappa)$ with $\protect\delta=2U\protect\kappa%
_2/\protect\kappa_1$ and $\protect\phi=\protect\pi$. The other parameters
are $\Delta_1=0$, $\protect\kappa_1=\protect\kappa$, $\protect\kappa_2=%
\protect\kappa/10$, $\protect\varepsilon_1=0.1\protect\kappa$, and $\protect%
\varepsilon_2=\protect\varepsilon_1\protect\sqrt{\protect\kappa_2/\protect%
\kappa_1}$.}
\label{fig6}
\end{figure*}

As an example, we show the second-order correlation $\log_{10}[g^{(2)}_{%
\mathrm{out}}(0)]$ in Fig.~\ref{fig6} for $\kappa _{2}=\kappa _{1}/10$. We
can see that the numerical results shown in Fig.~\ref{fig6} are very
similar to the Figs.~\ref{fig2} and \ref{fig3}, except that the optimal
detunings are $\delta\approx\pm 2U\kappa_2/\kappa_1$. In brief, we can
achieve scaling enhancement of photon blockade with $\kappa _{1} = \kappa
_{2}$ or $\kappa _{1} \neq \kappa _{2}$. In the main text, we set $%
\kappa_1=\kappa_2=\kappa$, without loss of generality.

\section{TLS-cavity interaction}

\label{AppC}

In this Appendix, we will derive the second-order correlation function of the
output field for the system consisting of two (uncoupled) cavities with a
TLS in one of them. The system can be described by a
JC model, including decay effects, as 
\begin{eqnarray}
H_{\mathrm{JC,eff}} &=&H_{\mathrm{JC}}-i\kappa _{1}a_{1}^{\dag
}a_{1}-i\kappa _{2}a_{2}^{\dag }a_{2}-i\frac{\kappa _{a}}{2}\sigma
_{+}\sigma _{-}  \notag \\
&=&\left( \Delta _{1}-i\kappa _{1}\right) a_{1}^{\dag }a_{1}+\left( \Delta
_{a}-i\frac{\kappa _{a}}{2}\right) \sigma _{+}\sigma _{-}  \notag \\
&&+g\left( a_{1}^{\dag }\sigma _{-}+\sigma _{+}a_{1}\right) +\left( \Delta
_{2}-i\kappa _{2}\right) a_{2}^{\dag }a_{2}  \notag \\
&&+i\varepsilon _{1}\left( a_{1}^{\dag }-a_{1}\right) +i\varepsilon
_{2}\left( a_{2}^{\dag }-a_{2}\right) ,
\end{eqnarray}%
and the wave function%
\begin{eqnarray}
\left\vert \varphi \right\rangle &=&C_{g00}\left\vert g,0,0\right\rangle
+C_{g10}\left\vert g,1,0\right\rangle +C_{g01}\left\vert g,0,1\right\rangle 
\notag \\
&&+C_{g20}\left\vert g,2,0\right\rangle +C_{g11}\left\vert
g,1,1\right\rangle +C_{g02}\left\vert g,0,2\right\rangle  \notag \\
&&+C_{e00}\left\vert e,0,0\right\rangle +C_{e10}\left\vert
e,1,0\right\rangle +C_{e01}\left\vert e,0,1\right\rangle ,
\end{eqnarray}%
with Fock-state basis truncated to the two-photon manifold, under the weak
driving conditions ($\varepsilon \ll \{\kappa _{1},\kappa _{2},\kappa _{a}\}$%
). Here, $\left\vert g,n_{1},n_{2}\right\rangle $ ($\left\vert
e,n_{1},n_{2}\right\rangle $) denotes the Fock state of $n_{1}$ photons in
cavity $a_{1}$, $n_{2}$ photons in cavity $a_{2}$, and the TLS in the ground
(excited) state, with the probability amplitude $C_{gn_{1}n_{2}}$ ($%
C_{en_{1}n_{2}}$). By substituting the wave function $\left\vert \varphi
\right\rangle $ and effective Hamiltonian $H_{\mathrm{JC,eff}}$ into the Schr%
\"{o}dinger equation, $d\left\vert \varphi \right\rangle /dt=-iH_{\mathrm{%
JC,eff}}\left\vert \varphi \right\rangle $, we get the dynamic equations for
the probability amplitudes $C_{gn_{1}n_{2}}$ ($C_{en_{1}n_{2}}$), and the
probability amplitudes $C_{gn_{1}n_{2}}$ ($C_{en_{1}n_{2}}$) can be obtained
analytically in the steady state $dC_{gn_{1}n_{2}}/dt=dC_{en_{1}n_{2}}/dt=0$%
. Under weak driving conditions, we have $\left\vert C_{g00}\right\vert
\approx 1\gg \left\{ \left\vert C_{g10}\right\vert ,\left\vert
C_{g01}\right\vert ,\left\vert C_{e00}\right\vert \right\} \gg \left\{
\left\vert C_{g20}\right\vert ,\left\vert C_{g11}\right\vert ,\left\vert
C_{g02}\right\vert ,\left\vert C_{e10}\right\vert ,\left\vert
C_{e01}\right\vert \right\} $, then the probability amplitudes of
one-particle excitation states ($C_{g10}$, $C_{g01}$, and $C_{e00}$) are
obtained as%
\begin{equation}
C_{g10}=\frac{-i\varepsilon \left( \Delta _{a}-i\frac{\kappa _{a}}{2}\right) 
}{\left( \Delta _{1}-i\kappa _{1}\right) \left( \Delta _{a}-i\frac{\kappa
_{a}}{2}\right) -g^{2}},
\end{equation}%
\begin{equation}
C_{e00}=\frac{i\varepsilon g}{\left( \Delta _{1}-i\kappa _{1}\right) \left(
\Delta _{a}-i\frac{\kappa _{a}}{2}\right) -g^{2}},
\end{equation}%
\begin{equation}
C_{g01}=\frac{-i\varepsilon }{\left( \Delta _{2}-i\kappa _{2}\right) },
\end{equation}%
and the probability amplitudes of two-photon states ($C_{g20}$, $C_{g11}$,
and $C_{g02}$) are obtained as 
\begin{widetext}
\begin{equation}
C_{g20}=-i\varepsilon \sqrt{2}\frac{\left( \Delta _{1}+\Delta _{a}-i\kappa
_{2}-i\kappa _{a}/2\right) C_{g10}-gC_{e00}}{\left( 2\Delta _{1}-i2\kappa
_{1}\right) \left( \Delta _{1}+\Delta _{a}-i\kappa _{2}-i\kappa
_{a}/2\right) -2g^{2}},
\end{equation}%
\begin{equation}
C_{g11}=\frac{-i\varepsilon \left[ \left( \Delta _{a}+\Delta _{2}-i\kappa
_{2}-i\kappa _{a}/2\right) C_{g10}-gC_{e00}\right] -i\varepsilon
C_{g01}\left( \Delta _{a}+\Delta _{2}-i\kappa _{2}-i\kappa _{a}/2\right) }{%
\left( \Delta _{1}+\Delta _{2}-i\kappa _{1}-i\kappa _{2}\right) \left(
\Delta _{a}+\Delta _{2}-i\kappa _{2}-i\kappa _{a}/2\right) -g^{2}},
\end{equation}%
\begin{equation}
C_{g02}=-\frac{i\sqrt{2}\varepsilon }{\left( 2\Delta _{2}-i2\kappa
_{2}\right) }C_{g01}.
\end{equation}%
The second-order correlation function for photons in the output field can be
given approximately by the probability amplitudes $C_{gn_{1}n_{2}}$ ($%
C_{en_{1}n_{2}}$) as 
\begin{eqnarray}
g_{\mathrm{out}}^{\left( 2\right) }\left( 0\right)  &\approx &\frac{1}{N_{%
\mathrm{out}}^{2}}\left[ 2\kappa _{1}^{2}\left\vert C_{g20}\right\vert
^{2}+2\kappa _{2}^{2}\left\vert C_{g02}\right\vert ^{2}+4\kappa _{1}\kappa
_{2}\left\vert C_{g11}\right\vert ^{2}+4\kappa _{1}\kappa _{2}\mathrm{Re}%
\left( e^{i2\phi }C_{g02}C_{g20}^{\ast }\right) \right.   \notag \\
&&\left. +4\sqrt{2}\kappa _{1}\sqrt{\kappa _{1}\kappa _{2}}\mathrm{Re}\left(
e^{i\phi }C_{g11}C_{g20}^{\ast }\right) +4\sqrt{2}\kappa _{2}\sqrt{\kappa
_{1}\kappa _{2}}\mathrm{Re}\left( e^{i\phi }C_{g02}C_{g11}^{\ast }\right) %
\right] 
\end{eqnarray}%
with $N_{\mathrm{out}}\approx \kappa _{1}\left\vert C_{g10}\right\vert
^{2}+\kappa _{2}\left\vert C_{g01}\right\vert ^{2}+2\sqrt{\kappa _{1}\kappa
_{2}}\mathrm{Re}\left( e^{i\phi }C_{g01}C_{g10}^{\ast }\right) $.
\end{widetext}

In order to obtain the parameter conditions for photon blockade enhancement
in the output field, based on the quantum interference between the
coefficients $C_{g20}$ and $C_{g11}$, the second-order correlation function
can be rewritten as%
\begin{eqnarray}
g_{\mathrm{out}}^{\left( 2\right) }\left( 0\right) &\approx &\frac{2\kappa
^{2}}{N_{\mathrm{out}}^{2}}\left\{ \left\vert C_{g20}+e^{i\phi }\sqrt{2}%
C_{g11}\right\vert ^{2}+\left\vert C_{g02}\right\vert ^{2}\right. \\
&&\left. +2\mathrm{Re}\left[ e^{i2\phi }\left( C_{g20}+e^{i\phi }\sqrt{2}%
C_{g11}\right) ^{\ast }C_{g02}\right] \right\}  \notag
\end{eqnarray}%
under the condition $\kappa _{1}=\kappa _{2}=\kappa _{a}/2=\kappa $. In
order to cancel the terms related to $C_{g20}$ and $C_{g11}$, we need%
\begin{equation}
\frac{|C_{g20}|}{|C_{g11}|}=\sqrt{2},
\end{equation}%
which is satisfied with 
\begin{equation}
\Delta _{2}=\pm \frac{2}{3}g,
\end{equation}%
under the optimal conditions ($\Delta _{1}=\Delta _{a}=\pm g$ and $g\gg
\kappa $) for photon blockade in cavity $a_{1}$. In the case of $\{\Delta
_{2}=-2g/3,\;\phi \approx \pi \}$ (or $\{\Delta _{2}=2g/3,\;\phi \approx 0\}$%
), the second-order correlation function for photons in the output field is
simplified as 
\begin{equation}
g_{\mathrm{out}}^{\left( 2\right) }\left( 0\right) \approx 16\left( \frac{%
\kappa }{g}\right) ^{4},
\end{equation}%
which is much smaller than the second-order correlation function for photons
in cavity $a_{1}$ as 
\begin{equation}
g_{\mathrm{1}}^{\left( 2\right) }\left( 0\right) =\frac{\left\langle
a_{1}^{\dag }a_{1}^{\dag }a_{1}a_{1}\right\rangle }{\left\langle a_{1}^{\dag
}a_{1}\right\rangle ^{2}}\approx \frac{2\left\vert C_{g20}\right\vert ^{2}}{%
\left\vert C_{g10}\right\vert ^{4}}\approx 36\left( \frac{\kappa }{g}\right)
^{2},
\end{equation}%
in the strong coupling regime $g\gg \kappa $.

\section{Second-order nonlinear interaction}

\label{AppD}

\begin{figure*}[tbp]
\centering
\includegraphics[bb=30 465 555 584, width=18 cm, clip]{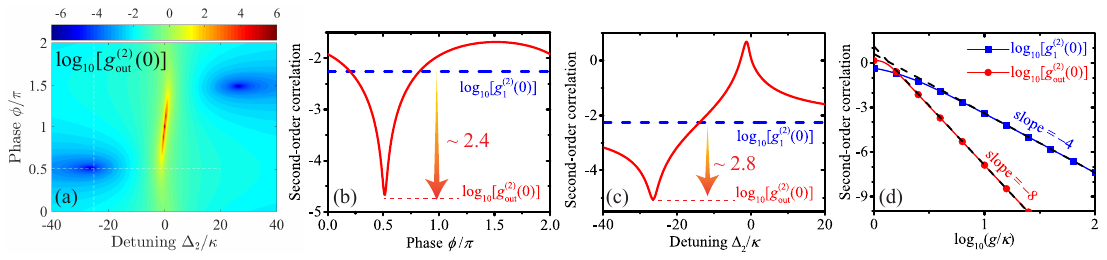}
\caption{(Color online) (a) The second-order correlation $\log_{10}[g^{(2)}_{%
\mathrm{out}}(0)]$ for different phase $\protect\phi/\protect\pi$ and
detuning $\Delta_2/\protect\kappa$ with $g=5\protect\kappa$. The
second-order correlations $\log_{10}[g^{(2)}_{\mathrm{out}}(0)]$ and $%
\log_{10}[g^{(2)}_{1}(0)]$ (b) versus phase $\protect\phi/\protect\pi$ with $%
\Delta_2=-g^2/\protect\kappa$ and $g=5\protect\kappa$, (c) versus detuning $%
\Delta_2/\protect\kappa$ with $\protect\phi=\protect\pi/2$ and $g=5\protect%
\kappa$, and (d) versus the second-order nonlinear interaction strength $%
\log_{10}(g/\protect\kappa)$ with $\Delta_2=-g^2/\protect\kappa$ and $%
\protect\phi=\protect\pi/2$. The other parameters are $\Delta_1=\Delta_b=0$, 
$\protect\kappa_b=2\protect\kappa$ and $\protect\varepsilon=0.001\protect%
\kappa$.}
\label{fig7}
\end{figure*}

The second-order nonlinear interaction in non-centrosymmetric materials is
another typical nonlinear interactions that has attracted great attentions
for it is usually orders of magnitude higher than the third-order
interaction~\cite{Zhang2019NaPho,Lu2020OPTICA}. In this Appendix, we will
consider the $\chi ^{(2)}$ interaction in one of the cavities, then the
system Hamiltonian becomes%
\begin{eqnarray}
H_{\mathrm{2nd}} &=&\Delta _{1}a_{1}^{\dag }a_{1}+\Delta _{b}b^{\dag
}b+g\left( a_{1}^{\dag 2}b+b^{\dag }a_{1}^{2}\right)+\Delta _{2}a_{2}^{\dag
}a_{2}  \notag \\
&&+i\varepsilon _{1}\left( a_{1}^{\dag }-a_{1}\right) +i\varepsilon
_{2}\left( a_{2}^{\dag }-a_{2}\right) .
\end{eqnarray}%
Here, $\Delta _{b}=\omega _{b}-2\omega _{p}$ is the detuning of the
second-harmonic mode ($b$ with frequency $\omega _{b}$), and $g$ is the
corresponding second-order nonlinear interaction strength.

In the presence of optical decay, the system can be written as an effective
Hamiltonian%
\begin{eqnarray}
H_{\mathrm{2nd,eff}} &=&H_{\mathrm{2nd}}-i\kappa _{1}a_{1}^{\dag
}a_{1}-i\kappa _{2}a_{2}^{\dag }a_{2}-i\kappa _{b}b^{\dag }b,  \notag \\
&=&\left( \Delta _{1}-i\kappa _{1}\right) a_{1}^{\dag }a_{1}+\left( \Delta
_{b}-i\kappa _{b}\right) b^{\dag }b  \notag \\
&&+g\left( a_{1}^{\dag 2}b+b^{\dag }a_{1}^{2}\right) +\left( \Delta
_{2}-i\kappa _{2}\right) a_{2}^{\dag }a_{2}  \notag \\
&&+i\varepsilon _{1}\left( a_{1}^{\dag }-a_{1}\right) +i\varepsilon
_{2}\left( a_{2}^{\dag }-a_{2}\right) ,
\end{eqnarray}%
where $\kappa _{b}$ is the one-sided decay rate of the second-harmonic mode $%
b$. For a very weak probe field, the state of the system can be truncated to
the first several Fock states as%
\begin{eqnarray}
\left\vert \psi ^{\prime }\right\rangle &=&C_{000}\left\vert
0,0,0\right\rangle +C_{010}\left\vert 0,1,0\right\rangle +C_{001}\left\vert
0,0,1\right\rangle  \notag \\
&&+C_{100}\left\vert 1,0,0\right\rangle +C_{020}\left\vert
0,2,0\right\rangle +C_{011}\left\vert 0,1,1\right\rangle  \notag \\
&&+C_{002}\left\vert 0,0,2\right\rangle .
\end{eqnarray}%
Here, $\left\vert n_{b},n_{1},n_{2}\right\rangle $ represents the Fock state
of $n_{b}$ photons in cavity $b$, $n_{1}$ photons in cavity $a_{1}$, and $%
n_{2}$ photons in cavity $a_{2}$, with the probability amplitude $%
C_{n_{b}n_{1}n_{2}}$. Substituting the wave function $\left\vert \psi
^{\prime }\right\rangle $ and effective Hamiltonian $H_{\mathrm{2nd,eff}}$
into the Schr\"{o}dinger's equation $i\partial \left\vert \psi ^{\prime
}\right\rangle /\partial t=H_{\mathrm{2nd,eff}}\left\vert \psi ^{\prime
}\right\rangle $, we get the dynamic equations for the probability
amplitudes $C_{n_{b}n_{1}n_{2}}$. For simplicity, we assume that $\kappa
_{b}=2\kappa _{1}=2\kappa _{2}=2\kappa $, $\Delta _{b}=\Delta _{1}=0$, $%
\varepsilon _{1}=\varepsilon _{2}=\varepsilon $. Under weak driving
conditions, i.e., $\varepsilon \ll \kappa $, we have $C_{000}\approx 1\gg
\left\{ \left\vert C_{010}\right\vert ,\left\vert C_{001}\right\vert
\right\} \gg \left\{ \left\vert C_{100}\right\vert ,\left\vert
C_{020}\right\vert ,\left\vert C_{011}\right\vert ,\left\vert
C_{002}\right\vert \right\} $, and the probability amplitudes in the steady
state are obtained analytically as%
\begin{equation}
C_{010}\approx \frac{\varepsilon }{\kappa },
\end{equation}%
\begin{equation}
C_{001}\approx \frac{-i\varepsilon }{\left( \Delta _{2}-i\kappa \right) },
\end{equation}%
for one-photon states, and%
\begin{equation}
C_{020}\approx \frac{\sqrt{2}\varepsilon ^{2}}{2\kappa ^{2}+g^{2}},
\end{equation}%
\begin{equation}
C_{011}\approx \left[ -\frac{i\varepsilon ^{2}}{\kappa }-\frac{\varepsilon
^{2}}{\left( \Delta _{2}-i\kappa \right) }\right] \frac{1}{\left( \Delta
_{2}-i2\kappa \right) },
\end{equation}%
\begin{equation}
C_{002}\approx -\frac{\sqrt{2}\varepsilon ^{2}}{2\left( \Delta _{2}-i\kappa
\right) \left( \Delta _{2}-i\kappa \right) },
\end{equation}%
for two-photon states.

Based on the probability amplitudes, the second-order correlation function
for photons in the output field can be rewritten as%
\begin{eqnarray}
g_{\mathrm{out}}^{\left( 2\right) }\left( 0\right) &\approx &\frac{2\kappa
^{2}}{N_{\mathrm{out}}^{2}}\left\{ \left\vert C_{020}+e^{i\phi }\sqrt{2}%
C_{011}\right\vert ^{2}+\left\vert C_{002}\right\vert ^{2}\right. \\
&&\left. +2\mathrm{Re}\left[ e^{i2\phi }\left( C_{020}+e^{i\phi }\sqrt{2}%
C_{011}\right) ^{\ast }C_{002}\right] \right\} .  \notag
\end{eqnarray}%
In order to realize destructive quantum interference between $C_{020}$ and $%
C_{011}$, i.e., $C_{010}+e^{i\phi }\sqrt{2}C_{001}\approx 0$, we have%
\begin{equation}
\Delta _{2}=\pm \frac{g^{2}}{\kappa },\qquad \phi =\mp \pi /2
\end{equation}%
for $\left\{ \left\vert \Delta _{2}\right\vert ,g\right\} \gg \kappa $.
Under the above conditions, the second-order correlation function for
photons in the output field is simplified as%
\begin{equation}
g_{\mathrm{out}}^{\left( 2\right) }\left( 0\right) \approx 13\left( \frac{%
\kappa }{g}\right) ^{8},
\end{equation}%
which is much smaller than the second-order correlation function for the
photons in the cavity $a_{1}$,%
\begin{equation}
g_{\mathrm{1}}^{\left( 2\right) }\left( 0\right) \approx 4\left( \frac{%
\kappa }{g}\right) ^{4}
\end{equation}%
in the strong coupling regime ($g>\kappa $).

Figure~\ref{fig7}(a) is a color plot of $\log_{10}[g^{(2)}_{\mathrm{out}%
}(0)]$ as a function of the phase $\phi/\pi$ and detuning $\Delta_2/\kappa$,
for $\Delta_1=\Delta_b=0$ and $g=5\kappa$. The minimum of $%
\log_{10}[g^{(2)}_{\mathrm{out}}(0)]$ is reached for $\phi\approx \pi/2$ and 
$\Delta_2\approx -g^2/\kappa$ (or $\phi\approx 3\pi/2$ and $\Delta_2\approx
g^2/\kappa$). Two cuts taken from the color plot for $\Delta_2\approx
-g^2/\kappa$ and $\phi=\pi/2$ are shown in Figs.~\ref{fig7}(b) and \ref%
{fig7}(c), respectively. The photon blockade is enhanced significantly as $%
g^{(2)}_{\mathrm{out}}(0)$ is about $2.4$ orders smaller than $%
g^{(2)}_{1}(0) $ at $\phi\approx \pi/2$ [Fig.~\ref{fig7}(b)] and about $2.8$
orders smaller at $\Delta_2\approx -g^2/\kappa$ [Fig.~\ref{fig7}(c)]. Both $%
\log_{10}[g^{(2)}_{\mathrm{out}}(0)]$ and $\log_{10}[g^{(2)}_{1}(0)]$ are
plotted as functions of $\log_{10}(g/\kappa)$ in Fig.~\ref{fig7}(d).
Different from the scaling behaviors for $\chi^{(3)}$ nonlinearity and
TLS-cavity interaction, in the strong second-order nonlinear regime $%
g/\kappa \gg 1$, the slope of $\log_{10}[g^{(2)}_{\mathrm{out}}(0)]$ versus $%
\log_{10}(g/\kappa)$ is $-8$, which is much larger than the slope of $-4$
for $\log_{10}[g^{(2)}_{1}(0)]$ versus $\log_{10}(g/\kappa)$. The numerical
results agree well with the analytical expressions in the strong nonlinear
regime [black dashed lines in Fig.~\ref{fig7}(d)]. Thus, the scheme we
proposed can change the scaling exponent of the second-order correlation on
the second-order nonlinear interaction strength from $-4$ to $-8$.

\section{Optomechanical interaction}

\label{AppE}

\begin{figure*}[tbp]
\includegraphics[bb=27 465 558 586, width=18 cm, clip]{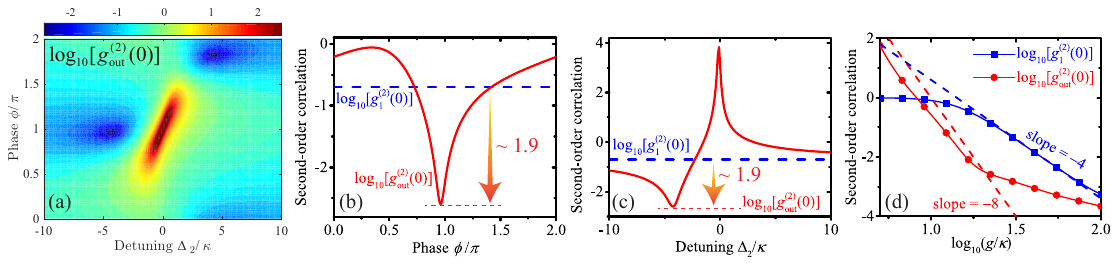}
\caption{(Color online) (a) The second-order correlation $\log_{10}[g^{(2)}_{%
\mathrm{out}}(0)]$ for different phase $\protect\phi/\protect\pi$ and
detuning $\Delta_2/\protect\kappa$ with $g=20\protect\kappa$. The
second-order correlations $\log_{10}[g^{(2)}_{\mathrm{out}}(0)]$ and $%
\log_{10}[g^{(2)}_{1}(0)]$ (b) versus phase $\protect\phi/\protect\pi$ with $%
\Delta_2=-4.3\protect\kappa$ and $g=20\protect\kappa$, (c) versus detuning $%
\Delta_2/\protect\kappa$ with $\protect\phi=0.96\protect\pi$ and $g=20%
\protect\kappa$, and (d) versus the optomechanical interaction strength $%
\log_{10}(g/\protect\kappa)$ with $\Delta_2=-2U_{\mathrm{om}}$ and $\protect%
\phi=0.96\protect\pi$. The other parameters are $\Delta_1=U_{\mathrm{om}}$, $%
\protect\omega_m=200\protect\kappa$, $\protect\gamma=0.01\protect\kappa$ and 
$\protect\varepsilon=0.01\protect\kappa$.}
\label{fig8}
\end{figure*}

In this Appendix, we consider a mechanical mode $c$ with frequency $\omega
_{m}$ in the cavity $a_{1}$ and they are coupled through optomechanical
interaction~\cite%
{Aspelmeyer2014RMP,Rabl2011PRL,Nunnenkamp2011PRL,Liao2013PRA,Kronwald2013PRA,XuXW2013PRA,Das2023PRL}%
, then the whole system is described by the Hamiltonian%
\begin{eqnarray}
H_{\mathrm{om}} &=&\left( \Delta _{1}-i\kappa _{1}\right) a_{1}^{\dag
}a_{1}+\left( \Delta _{2}-i\kappa _{2}\right) a_{2}^{\dag }a_{2}  \notag \\
&&+\left( \omega _{m}-i\gamma \right) c^{\dag }c+ga_{1}^{\dag }a_{1}\left(
c+c^{\dag }\right)  \notag \\
&&+i\varepsilon _{1}\left( a_{1}^{\dag }-a_{1}\right) +i\varepsilon
_{2}\left( a_{2}^{\dag }-a_{2}\right) ,
\end{eqnarray}%
where $\gamma $ is the mechanical decay rate and $g$ is the single-photon
optomechanical interaction strength. In order to understand the optimal
conditions for the strong photon blockade, it is convenient to transform the
Hamiltonian into a displaced oscillator representation $H_{\mathrm{om,eff}%
}=SH_{\mathrm{om}}S^{\dag }$, by the unitary transformation%
\begin{equation}
S=\exp \left[ \frac{g}{\omega _{m}}a_{1}^{\dag }a_{1}\left( c^{\dag
}-c\right) \right] ,
\end{equation}%
as 
\begin{eqnarray}
H_{\mathrm{om,eff}} &\approx &\left( \Delta _{1}^{\prime }-i\kappa
_{1}\right) a_{1}^{\dag }a_{1}+\left( \Delta _{2}-i\kappa _{2}\right)
a_{2}^{\dag }a_{2}  \notag \\
&&+\left( \omega _{m}-i\gamma \right) c^{\dag }c-U_{\mathrm{om}}a_{1}^{\dag
}a_{1}^{\dag }a_{1}a_{1}  \notag \\
&&+i\varepsilon _{1}\left( a_{1}^{\dag }-a_{1}\right) +i\varepsilon
_{2}\left( a_{2}^{\dag }-a_{2}\right) .  \label{D_OM}
\end{eqnarray}%
Here $\Delta _{1}^{\prime }\equiv \Delta _{1}-U_{\mathrm{om}}$, $U_{\mathrm{%
om}}\equiv g^{2}/\omega _{m}$, and $i\varepsilon _{1}\left\{ a_{1}^{\dag
}\exp \left[ \frac{g}{\omega _{m}}\left( c^{\dag }-c\right) \right] -\mathrm{%
H.c.}\right\} \approx i\varepsilon _{1}\left( a_{1}^{\dag }-\mathrm{H.c.}%
\right) $ for $g<\omega _{m}$. We also assume $\kappa _{1}=\kappa
_{2}=\kappa \gg \gamma $, $\varepsilon _{1}=\varepsilon _{2}=\varepsilon $, $%
\varepsilon \ll \kappa $, in the following.

The effective Hamiltonian~(\ref{D_OM}) for the system with optomechanical
interaction is almost the same as the one~(\ref{3thr}) for the system with $%
\chi ^{(3)}$ nonlinearity, except the phonon mode. According to the results
obtained in Appendix~\ref{AppB}, we can obtain the second-order correlation%
\begin{equation}
g_{\mathrm{out}}^{\left( 2\right) }\left( 0\right) \approx \frac{1}{16}%
\left( \frac{\kappa }{U_{\mathrm{om}}}\right) ^{4}=\frac{1}{16}\left( \frac{%
\kappa \omega _{m}}{g^2}\right) ^{4}
\end{equation}%
for photons in the output field, and 
\begin{equation}
g_{\mathrm{1}}^{\left( 2\right) }\left( 0\right) \approx \left( \frac{\kappa 
}{U_{\mathrm{om}}}\right) ^{2}=\left( \frac{\kappa \omega _{m}}{g^2} \right)
^{2}
\end{equation}%
for photons in the cavity $a_{1}$, with the optimal conditions 
\begin{equation}
\Delta _{1}=U_{\mathrm{om}},\qquad \Delta _{2}=-2U_{\mathrm{om}},\qquad \phi
\approx \pi ,
\end{equation}%
or 
\begin{equation}
\Delta _{1}=U_{\mathrm{om}},\qquad \Delta _{2}=2U_{\mathrm{om}},\qquad \phi
\approx 0.
\end{equation}

We show $\log_{10}[g^{(2)}_{\mathrm{out}}(0)]$ numerically in Fig.~\ref%
{fig8}. We can see that the photon blockade is enhanced significantly as $%
g^{(2)}_{\mathrm{out}}(0)$ is about $1.9$ orders smaller than $%
g^{(2)}_{1}(0) $ in Figs.~\ref{fig7}(b) and \ref{fig7}(c). Both $%
\log_{10}[g^{(2)}_{\mathrm{out}}(0)]$ and $\log_{10}[g^{(2)}_{1}(0)]$ are
plotted as functions of $\log_{10}(g/\kappa)$ in Fig.~\ref{fig8}(d).
Similar to the scaling behaviors for second-order nonlinear interaction, in
the strong optomechanical interaction regime $g/\kappa \gg 1$, the slope of $%
\log_{10}[g^{(2)}_{\mathrm{out}}(0)]$ versus $\log_{10}(g/\kappa)$ is $-8$,
which is much larger than the slope of $-4$ for $\log_{10}[g^{(2)}_{1}(0)]$
versus $\log_{10}(g/\kappa)$. But the numerical result is a sharp departure
from the analytical expressions for the output field [red dashed lines in
Fig.~\ref{fig8}(d)]. That is because there are many phonon states in the
optomechanical system, and the resonant transition between the states with
different phonons may suppress photon blockade~\cite{Liao2013PRA,XuXW2013PRA}%
.

\bibliography{ref}


\end{document}